# Structural, elastic, electronic, and bonding properties of intermetallic $Nb_3Pt$ and $Nb_3Os$ compounds: a DFT study

**M. I. Naher[a], F. Parvin[a], A. K. M. A. Islam[b], S. H. Naqib[a*]**

[a]Department of Physics, University of Rajshahi, Rajshahi 6205, Bangladesh
[b]International Islamic University Chittagong, 154/A College Road, Chittagong-4203, Bangladesh

*Corresponding author: salehnaqib@yahoo.com

## Abstract

Theoretical investigation of structural, elastic, electronic and bonding properties of *A*-15 Nb-based intermetallic compounds $Nb_3B$ (*B* = Pt, Os) have been performed using first principles calculations based on the density functional theory (DFT). Optimized cell parameters are found to be in good agreement with available experimental and theoretical results. The elastic constants at zero pressure and temperature are calculated and the anisotropic behaviors of the compounds are studied. Both the compounds are mechanically stable and ductile in nature. Other elastic parameters such as Pugh's ratio, Cauchy pressure, machinability index are derived for the first time. $Nb_3Os$ is expected to have good lubricating properties compared to $Nb_3Pt$. The electronic band structure and energy density of states (DOS) have been studied with and without spin-orbit coupling (SOC). The band structures of both the compounds are spin symmetric. Electronic band structure and DOS reveal that both the compounds are metallic and the conductivity mainly arises from the Nb 4*d* states. The Fermi surface features have been studied for the first time. The Fermi surfaces of $Nb_3B$ contain both hole- and electron-like sheets which change as one replaces Pt with Os. The electronic charge density distribution shows that $Nb_3Pt$ and $Nb_3Os$ both have a mixture of ionic and covalent bonding. The charge transfer between atomic species in these compounds has been explained by the Mulliken bond population analysis.

## 1. Introduction

In the advancement of many technologically important science and engineering branches, intermetallic compounds (IMCs) have played a significant role. Intermetallic compounds are one of the oldest and most important man made materials, being subject of constant interest for physicists, chemists and materials scientists [1-8]. Many intermetallic compounds display attractive combination of physical and mechanical properties, including high melting point, low density and good oxidation or corrosion resistance. They have vast applications in aerospace industry, aircraft, automotive engine, biomedical instrumentations, microelectronics, electronics, batteries, hydrogen storage systems, and chemical industries. In every solder joint, between the solder and the system of interest, a layer is present containing one or more intermetallic compounds [9-11].

A great deal of effort has been made in the last few decades to raise the critical temperature $T_c$ of superconductor compounds. It is well known that the compounds with A-15 type structure are known to exhibit superconductivity and physical properties of these compounds influence the superconducting parameters. In this context the A-15 family of $A_3B$ compounds has been extensively studied theoretically and experimentally [12-26] and comprehensive reviews of its interesting properties were published [27-28]. Even at normal state, the studies of the physical properties of these compounds are interesting [29-31]. Intermetallic compounds $Nb_3Pt$ and $Nb_3Os$ shows superconductivity. Also the studies of X-ray photoemission spectra of the A-15 type compounds $Nb_3B$ (Pt, Os) have indicated that the energy bands of Nb $4d$ and $5d$ orbital of these $A_3B$ compounds appear to be more and more separated with increasing atomic number of the $B$ element [32–35]. For the above reasons, physical properties of these compounds such as structural, elastic, electronic, bonding, optical, thermal, etc., are quite important to know. To the best of our knowledge, the elastic properties (except bulk modulus), spin polarized electronic properties (band structure, density of states, Fermi surface and charge density), Mulliken bond population and hardness of both $Nb_3Pt$ and $Nb_3Os$ have not yet been studied theoretically or experimentally, yet. Therefore, we undertook this project to explore some of these properties in details in this study.

We are interested to calculate the mechanical properties because they are important for different application related physical properties. They are also related to different fundamental solid state and thermal properties. The electronic properties such as band structure, density of states, Fermi surface and charge density are related to charge transport, electronic and thermal processes. Mulliken bond population analysis elucidates the bonding nature of the compounds.

This paper is organized as follows. In Section 2, computational scheme is described in brief. Section 3 consists of results and analysis of all the physical properties under study. Finally, in Section 4 we have discussed the theoretical results and drawn the main conclusions of this study.

## 2. Computational Methods

CASTEP (Cambridge Serial Total Energy Package) code [36] has been used to explore various properties of $Nb_3Pt$ and $Nb_3Os$. This code is based on the Density functional theory (DFT), which uses a total energy plane wave pseudopotential method [37-38]. The interaction between the valence electrons and ion cores has been represented by the Vanderbilt-type ultrasoft pseudo-potential for Nb, Pt and Os atoms [39]. The exchange-correlation energy was calculated using Generalized Gradient Approximation (GGA) of the Perdew–Burke–Ernzerhof (PBE) scheme [40]. The valence electron configurations used in this research were $4s^2\ 4p^6\ 4d^4\ 5s^1$ for Nb, $5d^9,\ 6s^1$ for Pt and $5s^2\ 5p^6\ 5d^6\ 6s^2$ for Os, respectively. The Brillouin zone (BZ) integrations were performed using the Monkhorst and Pack [41] *k*-point meshes. For geometry optimization for both $Nb_3Pt$ and $Nb_3Os$ 16×16×16 (120 *k*-points) *k*-mesh has been used to integrate the wave function in the BZ. The cut-off energy of 500 eV and 400 eV has been used for the $Nb_3Pt$ and $Nb_3Os$, respectively. The Fermi surfaces were obtained by sampling the whole BZ with the *k*-point meshes 25×25×25. These parameters are sufficient for well converged total energy, geometrical configurations and elastic moduli. The magnetic ground-state properties and electronic properties were studied using relaxed unit cell parameter and optimized with 120 *k*-points with the 16×16×16 Monkhorst-Pack grid and cut-off energy 500 eV for $Nb_3Pt$ and 400 eV for $Nb_3Os$.

The geometry optimization of $Nb_3Pt$ and $Nb_3Os$ were done using Broyden–Fletcher–Goldfarb–Shanno (BFGS) minimization scheme [42]. Optimization is performed using convergence thresholds of $10^{-5}$ eV/atom for the total energy and $10^{-3}$ Å for maximum displacement. Maximum stress and force were 0.05 GPa and 0.03 eV/Å, respectively for all calculations. The elastic constants were determined by applying a set of homogenous deformations with a finite value and calculating the resulting stress [43]. The elastic stiffness coefficients were determined from a linear fit of the calculated stress as a function of strain.

After optimization, the elastic constants, band structure, density of states, charge density, Mulliken population were calculated. For all the equilibrium structures, the Mulliken populations were investigated using a projection of the plane-wave states onto a linear combination of atomic orbital basis sets, [44-45] which is widely used to perform charge transfers and population analysis.

## 3. Physical Properties of $Nb_3Pt$ and $Nb_3Os$

### 3.1. *Structural Properties*

Niobium based A-15 intermetallic $Nb_3B$ ($B$ = Pt and Os) crystallize in the cubic structure with space group *Pm-3n* (no. 223), consisting of six *A* atoms which lie on the surface and form chains along the axis directions and two *B* atoms per unit cell which occupy the bcc sites in the unit cell. The Nb and *B* atoms occupy the following Wyckoff positions in the unit cell [46-47], Nb atoms: (0.25, 0, 0.5); (0.5, 0.25, 0); (0, 0.5, 0.25); (0.75, 0, 0.5); (0.5, 0.75, 0); (0, 0.5, 0.75) and *B* atoms: (0, 0, 0); (0.5, 0.5, 0.5). Fig.1. shows the equilibrium crystal structures of these compounds. The unit cell consists of two formula units (Z = 2) and 8 atoms. Table 1 summarized the results of first-principle calculations of the structural properties without spin consideration of these compounds, together with available experimental and theoretical values [48-50] for comparison. Our calculated values of lattice parameter are 0.77% and 0.38% larger than experimental values and 0.19% and 0.39% smaller from theoretical values for $Nb_3Pt$ and $Nb_3Os$, respectively. This comparison shows that our results are in very good agreement with both theoretical and experimental results. Since the atomic radius of Pt ($r$ = 1.83 Å) is larger than Os ($r$ = 0.63 Å), the lattice constant for $Nb_3Pt$ should be larger than $Nb_3Os$ with the same crystal structure. The optimized cell parameters of $Nb_3Pt$ and $Nb_3Os$ for ground-state spin configuration shows no deviation from the estimations without SOC. Therefore, as far as the structural parameters are concerned, SOC plays no significant role.

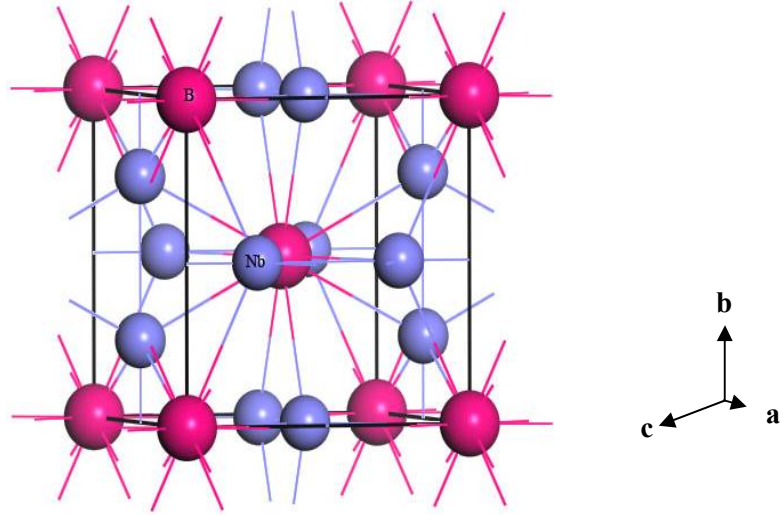

**Figure 1.** Crystal structure of Nb₃*B* (*B* = Pt, Os) unit cell.

**Table 1**

Calculated (without SOC) and experimental lattice constants $a$ (Å), equilibrium volume $V_o$ (Å³), bulk modulus $B$ (GPa) of Nb₃Pt and Nb₃Os.

| Compounds | $a$ | $V_o$ | $B$ | Ref. |
|---|---|---|---|---|
| Nb₃Pt | 5.19 | 139.74 | 199.96 | This |
| | 5.15 | - | - | [48,49][Exp.] |
| | 5.20 | - | 201.76 | [50][Theo.] |
| Nb₃Os | 5.16 | 137.20 | 225.26 | This |
| | 5.14 | - | - | [48,49][Exp.] |
| | 5.18 | - | 218.78 | [50][Theo.] |

### *3.2. Mechanical Properties*

Elastic constants of solids provide with a link between the mechanical and dynamic behavior of crystal and give important information about the response of a crystal to the external force as characterized by the bulk modulus $B$, shear modulus $G$, Young's modulus $Y$ and Poisson's ratio, ν. They play an important role in determining the strength of the material. They also provide information regarding mechanical stability and stiffness of the system. For the material with cubic symmetry, there are only three independent single crystalline elastic constants: $C_{11}$, $C_{12}$, and $C_{44}$. These calculated elastic constants are tabulated in Table 2. The requirement of mechanical stability in a cubic structure leads to the following conditions [51]: $C_{11}+2C_{12} > 0$, $C_{44} > 0$, $C_{11}-C_{12} > 0$. As shown in Table 2, both the compounds satisfy these criteria. It indicates that these compounds are mechanically stable. The tetragonal shear modulus, $C'$, is related to $C_{11}$ and $C_{12}$ by the following relation: $C' = (C_{11}- C_{12})/2$.

The polycrystalline elastic moduli can be calculated from the single crystalline elastic constants. Calculated values of bulk modulus, $B$ and shear modulus, $G$ are given in Table 3. To get the Young's modulus ($Y$), Poisson's ratio (v), and shear anisotropy factor ($A$) at zero pressure are, following equations are used [52]:

$$Y = \frac{9BG}{(3B + G)} \tag{1}$$

$$v = \frac{(3B - 2G)}{2(3B + G)} \tag{2}$$

and $$A = \frac{2C_{44}}{(C_{11} - C_{12})} \tag{3}$$

We observe that, $C_{44}$, which reflects the resistance to share deformation, is lower than $C_{11}$, which is related to the unidirectional compression along the principle crystallographic directions. This means that the cubic cell is more easily deformed by a share in comparison to the unidirectional compression. There is a correlation between the elastic constants and the melting temperature of a solid [53-54]. By comparing the elastic constants one can say that the melting point of Nb$_3$Os should be larger than that of Nb$_3$Pt.

Estimated mechanical parameters are presented in Table 3. The calculated result shows that the bulk modulus of Nb$_3$Pt is smaller than Nb$_3$Os, which agrees with the earlier theoretical results [50]. For both Nb$_3$Pt and Nb$_3$Os, $B > G$, which indicates that the shear modulus is the parameter that limits the mechanical stability. Young's modulus of Nb$_3$Os is larger than that of Nb$_3$Pt, indicating that Nb$_3$Os is stiffer than Nb$_3$Pt. The Poisson's ratio can illustrate the bonding properties of materials. Poisson's ratio is different, for different dominating bonding [55]. As for covalent materials, the value of v is small (typically v = 0.10); whereas for ionic materials, the typical value of v is 0.25; for metallic materials, v is typically 0.33 [55]. The values of v for Nb$_3$Pt and Nb$_3$Os are 0.31 and 0.32, respectively. This suggests that both the compounds should show metallic behavior. Since these materials predicted to be metallic, conduction of heat is expected to be predominantly by electrons at low temperatures. For a cubic material with its two share constants, $C_{44}$ and the apparent tetragonal shear modulus $C'$, the values of $G$ should be between these two components. The smaller the $C'$, the lower is the Young's modulus [56-57].

Some of the factors on which the brittle and ductile nature of a given material depends are as follows: Pugh's index ($G/B$), Poisson's ratio (v) and Cauchy pressure ($C_{12} - C_{44}$). Pugh [58-59] proposed an empirical relationship to distinguish the physical properties (ductility and brittleness) of materials. A material is ductile if the $G/B$ is less than 0.5. Otherwise it is brittle. In our case, for both the compounds, $G/B$ is much less than 0.5, i.e., these compounds should behave in a ductile manner. $Nb_3Os$ is more ductile than $Nb_3Pt$. Poisson's ratio is also an indicator of brittle/ductile behavior. For brittle materials $\nu \sim 0.1$, whereas for ductile metallic material $\nu$ is typically 0.33 [55]. Here, the Poisson's ratio of $Nb_3Pt$ and $Nb_3Os$ are 0.31 and 0.32, respectively, which again suggests that both the compounds should behave in a ductile manner. On the other hand, a positive Cauchy pressure suggests ductility for a material, whereas a negative value suggests brittleness [60]. The calculated values of Cauchy pressure for both $Nb_3Pt$ and $Nb_3Os$ are tabulated in Table 2. The Cauchy pressure of both the compounds is positive, which also indicates that both these compounds are ductile in nature.

The machinability index $\mu_M$, due to Sun *et al.* [61], is defined as,

$$\mu_M = B/C_{44}$$

which may be interpreted as a measure of plasticity [62]. Large value of $B/C_{44}$ indicates that the corresponding material has excellent lubricating properties. The $B/C_{44}$ values for $Nb_3Pt$ and $Nb_3Os$ are 3.15 and 3.35, respectively. $B/C_{44}$ values for $Nb_3Os$ is comparable to that for gold ($B/C_{44}$ = 4.17) [63]. Since gold has wonderful lubricating properties [64], $Nb_3Os$ is expected to have good lubricating properties. Since $B$ is a weighted average of $C_{11}$ and $C_{12}$ and stability requires that $C_{12}$ be smaller than $C_{11}$, we are then left with the result that $B$ is required to be intermediate in value between $C_{11}$ and $C_{12}$: $C_{12} < B < C_{11}$ [65-66].

**Table 2**

The calculated elastic constants, $C_{ij}$ (GPa), Cauchy pressure, $(C_{12} - C_{44})$ (GPa) and tetragonal shear modulus, $C'$ (GPa) for $Nb_3Pt$ and $Nb_3Os$ at $P$ = 0 GPa and $T$ = 0 K.

| Compounds | $C_{11}$ | $C_{12}$ | $C_{44}$ | $(C_{12}-C_{44})$ | $C'$ |
|-----------|----------|----------|----------|-------------------|------|
| $Nb_3Pt$ | 373.58 | 113.15 | 63.57 | 49.58 | 130.21 |
| $Nb_3Os$ | 413.09 | 131.34 | 67.26 | 64.07 | 140.38 |

Shear anisotropy factor ($A$) is an important parameter that is related to the structural stability, defect dynamics and elastic anisotropy of crystals [67]. For isotropic crystal, $A$ = 1, while any value smaller or larger than 1 points towards an elastic anisotropy. The magnitude of deviation from 1 is a measure of the degree of elastic anisotropy possessed by the crystal. The calculated values of $A$ for $Nb_3Pt$ and $Nb_3Os$ are 0.49 and 0.48, respectively, implying significant anisotropy.

**Table 3**

The calculated bulk modulus $B$ (in GPa), shear modulus $G$ (in GPa), Young's modulus $Y$ (in GPa), Pugh's indicator $G/B$, machinability index $B/C_{44}$, Poisson's ratio $v$, anisotropy factor $A$ for $Nb_3Pt$ and $Nb_3Os$ at $P = 0$ GPa and $T = 0$ K.

| Compounds | $B$ | $G$ | $Y$ | $G/B$ | $B/C_{44}$ | $v$ | $A$ | Ref. |
|-----------|-----|-----|-----|-------|-----------|-----|-----|------|
| $Nb_3Pt$ | 199.96 | 85.08 | 223.54 | 0.43 | 3.15 | 0.31 | 0.49 | This |
| | 201.76 | - | - | - | - | - | - | [50][Theo.] |
| $Nb_3Os$ | 225.26 | 90.87 | 240.31 | 0.40 | 3.35 | 0.32 | 0.48 | This |
| | 218.78 | - | - | - | - | - | - | [50][Theo.] |

### 3.3. Debye temperature

Debye temperature is an important fundamental thermodynamic parameter for solids. It leads to estimates of many physical properties such as melting temperature, specific heat, lattice vibration, thermal conductivity, and thermal expansion. It also sets the energy scale for electron-phonon interaction in superconductors. The Debye temperature has also relation with the vacancy formation energy in metals. Reliable estimates of Debye temperature for different types of materials can be obtained from the elastic moduli [68–70]. The average elastic wave velocity $v_a$ in a crystal is given by

$$v_a = \left[ \frac{1}{3} \left( \frac{2}{v_t^3} + \frac{1}{v_l^3} \right) \right]^{-\frac{1}{3}} \qquad (4)$$

where, $v_t$ and $v_l$ are the transverse and longitudinal wave velocities, respectively. The transverse velocity $v_t$ is expressed as

$$v_t = \sqrt{\frac{G}{\rho}} \qquad (5)$$

where, G is the shear modulus and $\rho$ is the density. The longitudinal wave velocity $v_l$ is obtained from

$$v_l = \sqrt{\frac{B + 4G/3}{\rho}} \qquad (6)$$

The Debye temperature, $\Theta_D$, can now be expressed as [71]

$$\Theta_D = \frac{h}{k_B}\left(\frac{3n}{4\pi V_0}\right)^{1/3} v_a \qquad (7)$$

where, $h$ is Planck's constant, $k_B$ is the Boltzmann's constant, $V_0$ is the volume of unit cell and $n$ is the number of atoms in unit cell.

The calculated Debye temperature $\Theta_D$ along with sound velocities $v_l$, $v_t$, and $v_a$ are presented in Table 4.

**Table 4**

Density $\rho$ (in g/cm$^3$), transverse $v_t$ (in ms$^{-1}$), longitudinal $v_l$ (in ms$^{-1}$), average elastic wave $v_a$ velocities (in ms$^{-1}$), and Debye temperature $\Theta_D$ (K) for Nb$_3$Pt and Nb$_3$Os.

| Compounds | $P$ | $v_t$ | $v_l$ | $v_a$ | $\Theta_D$ | Ref. |
|---|---|---|---|---|---|---|
| Nb$_3$Pt | 10.228 | 2884.2 | 5535.5 | 2976.6 | 218.98 | This |
| | - | - | - | - | 242 | [72][Exp.] |
| Nb$_3$Os | 10.311 | 2968.7 | 5796.3 | 3066.3 | 226.90 | This |
| | - | - | - | - | 207 | [72][Exp.] |

### 3.4. Electronic density of states and band structure

For Nb$_3$Pt and Nb$_3$Os, the calculated total and partial (projected) density of states (TDOSs and PDOSs, respectively), as a function of energy, $(E - E_F)$ are presented in Figs. 2(a) and (b), respectively. The vertical broken line denotes the Fermi level, $E_F$. To understand the contribution of each atom to the total TDOSs, we have calculated the PDOSs of Nb, Pt and Os in Nb$_3$Pt and Nb$_3$Os. The non-zero values of total DOSs at the Fermi level is the evidence that both Nb$_3$Pt and Nb$_3$Os should exhibit metallic conductivity. At the Fermi energy ($E_F$), the values of TDOSs for Nb$_3$Pt and Nb$_3$Os are 9.77 and 5.20 states per eV per unit cell, respectively. Hence we can say that Nb$_3$Pt is more conducting than Nb$_3$Os. Near the Fermi level the main contribution on TDOSs comes from Nb 4$d$ orbitals. Thus, the electrical conductivity of both the compounds is mainly dominated by Nb 4$d$ states. The chemical and mechanical stability of both Nb$_3$Pt and Nb$_3$Os are also mainly affected by the properties of Nb 4$d$ bonding electronic states.

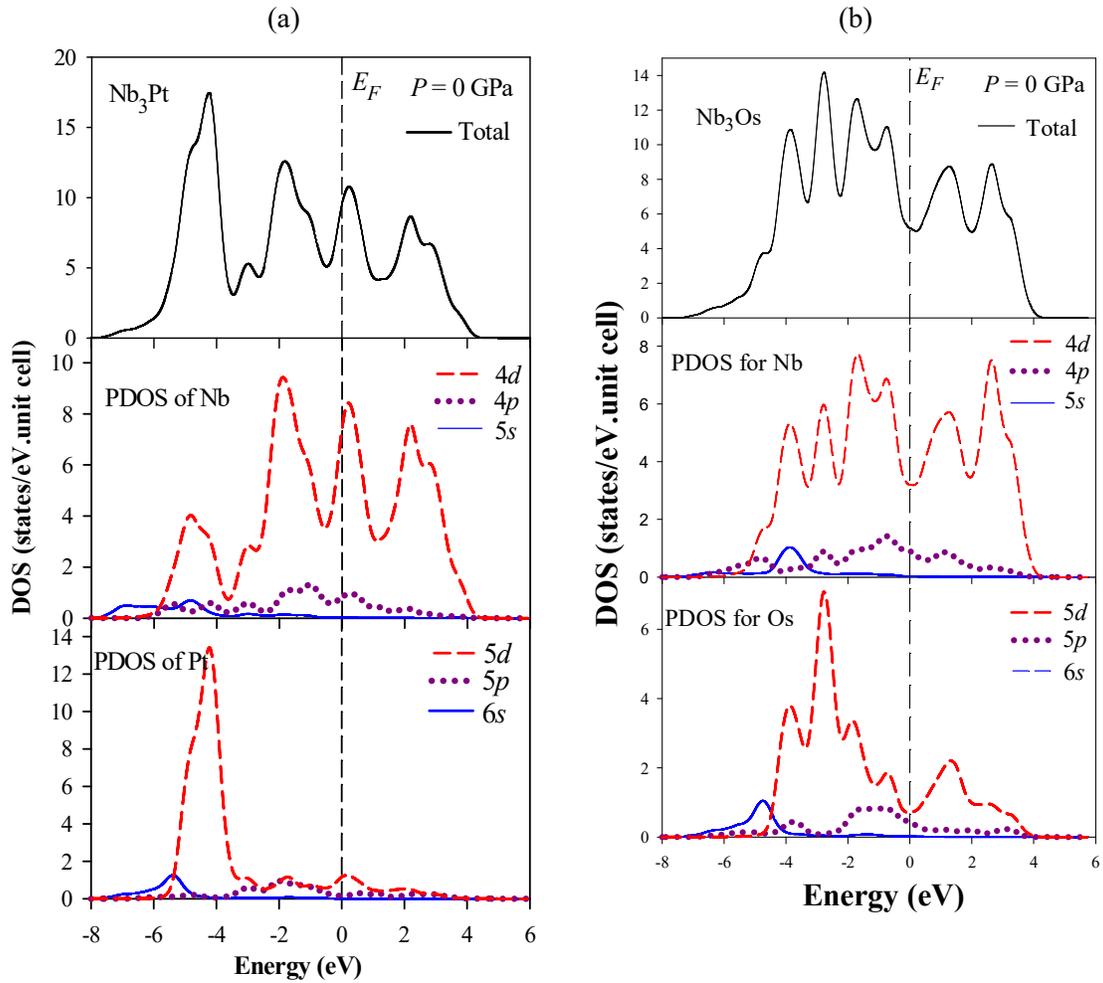

**Figure 2.** Total and partial density of states (DOS) of (a) $Nb_3Pt$ and (b) $Nb_3Os$ as a function of energy. The Fermi level is placed at the origin.

From the TDOSs curves of $Nb_3Pt$ and $Nb_3Os$, we can see that the valence and conduction bands of $Nb_3Pt$ is located between -7.84 and 4.47 eV, which is mainly dominated by Nb $4d$ orbitals with small contribution of $4p$ states of Nb and $5p$, $5d$ states of Pt. On the other hand, the valence and conduction bands of $Nb_3Os$ is located between -7.34 to 4.18 eV, which is dominated by $4d$ and $4p$ states of Nb and with small contribution of $5p$ and $5d$ orbitals of Os. It is noticed that Nb $5s$ and Pt/Os $6s$ states do not contribute to the DOSs at the Fermi level.

Since both $Nb_3Pt$ and $Nb_3Os$ are metallic system, the DOSs at Fermi level, $N(E_F)$, is a key parameter for the electronic stability purpose. The phase stability of intermetallic compounds depends on the location of Fermi level and the value of $N(E_F)$ [73-74]. Systems with lower values of $N(E_F)$ are more stable than those with higher values of $N(E_F)$. The total DOSs of $Nb_3Pt$ and Nb₃Os at the Fermi level is 9.77 and 5.20 states per eV per unit cell, respectively. Thus, Nb₃Os is electronically more stable than Nb₃Pt.

Another important feature is the existence of a pseudogap or quasi-gap in the DOSs in the vicinity of the Fermi level, which is also related to electronic stability [75]. The formation of pseudogap has no common understanding so far. The ionic origin and hybridization origin are mainly responsible for pseudogap formation [76]. The pseudogap separates bonding states from nonbonding/antibonding states. In both the cases, the Fermi level lies to the right of the pseudogap i.e., at the antibonding regions/ states.

TDOSs around the Fermi level of Nb₃Os have a number of peaks, which indicate that a dramatic change in physical properties may occur if the Fermi level is shifted by doping, stress, etc. It can be seen from TDOSs curves that there are 2 and 4 bonding peaks below the Fermi level for Nb₃Pt and Nb₃Os, respectively. For Nb₃Pt, strong bonding peak (at -4.27 eV) is formed by the hybridization between Pt 5$d$ orbitals and Nb 4$s$ orbitals. Similarly, for Nb₃Os a strong peak (at -2.78 eV) is formed due to hybridization between Os 5$d$ and Nb 4$d$ orbitals.

We have neglected the spin-orbit coupling (SOC) for the DOS calculation presented upto this point. The spin-polarized total density of states (TDOSs) of the intermetallic compounds for both spin-up and -down are shown in Figs. 3. The gross features are quite similar to those obtained without SOC. The spin polarization at the Fermi level (or Fermi energy $E_F$) is related to the spin dependent DOS via the expression [77]

$$P = \frac{n \uparrow (E_F) - n \downarrow (E_F)}{n \uparrow (E_F) + n \downarrow (E_F)}$$

which should be 100% in the completely spin asymmetric case.

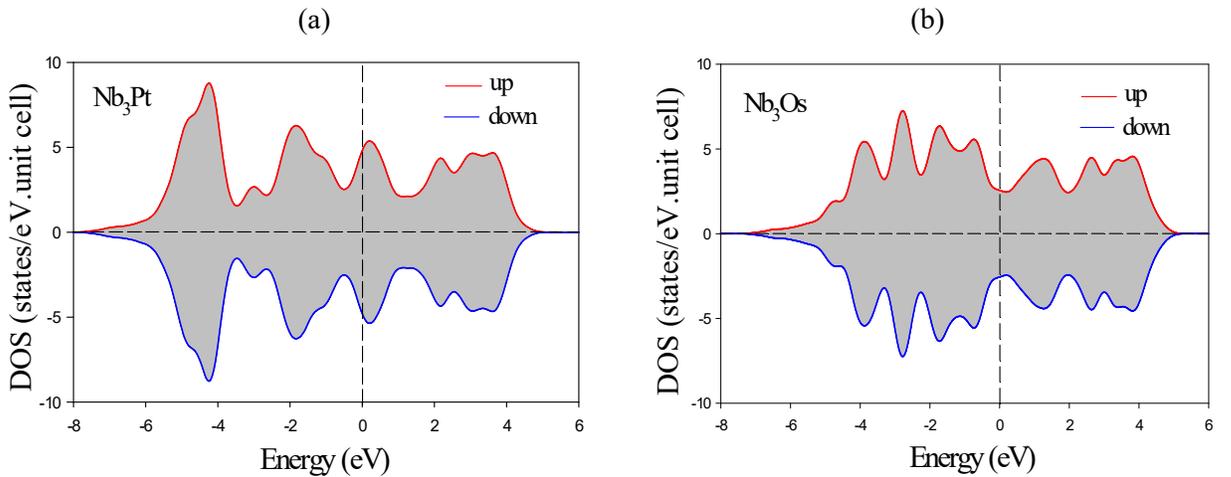

**Figure 3.** Total density of states (DOS) for spin up and down configuration of (a) Nb₃Pt and (b) Nb₃Os.

It is seen from Fig. 3 that there is negligible spin asymmetry in the TDOSs of $Nb_3Pt$ and $Nb_3Os$ intermetallic compounds.

We have calculated the electronic band structures of $Nb_3Pt$ and $Nb_3Os$, without and with SOC consideration, along several high symmetry directions ($R - \Gamma - X - M - \Gamma$) in the first Brillouin zone. Figs. 4 show the band structures without SOC of the compounds at zero pressure. The Valence Band (VB) and Conduction Band (CB) are considerably overlapped and there is no band gap at the Fermi level. Thus both $Nb_3Pt$ and $Nb_3Os$ should exhibit metallic behavior.

The number of bands, as seen from Fig. 4, (in the first Brillouin zone) for $Nb_3Pt$ and $Nb_3Os$ are 59 and 67, respectively. The bands which cross the Fermi energy are shown (colored) in Figs 4(a) and 4(b) indicating their band number. The numbers of bands that cross the Fermi energy are 49, 50 and 51 for $Nb_3Pt$ and 54, 55, 56 and 57 for $Nb_3Os$. These bands are both electron- and hole-like, thus resulting in a multiband system dominated by Nb-$d$ electronic states. Specifically hole-like features are observed at symmetry points $R$, $\Gamma$, and midway the $R$ - $\Gamma$ line for $Nb_3Pt$ and $M$, $R$, $X$, midway the $X$ - $M$ and $M$ - $\Gamma$ line for $Nb_3Os$. On the other hand, electron-like features are observed at the symmetry points $\Gamma$, $X$ and $M$ for $Nb_3Pt$ and $\Gamma$, $R$, and $X$ for $Nb_3Os$. In both $Nb_3Pt$ and $Nb_3Os$, the lowest energy bands are formed from $s$, $p$, and $d$ states of Pt and Os atoms. The band lying in the range of -3.62 to 4.36 eV for $Nb_3Pt$ mainly arises from 4$d$ states of Nb.

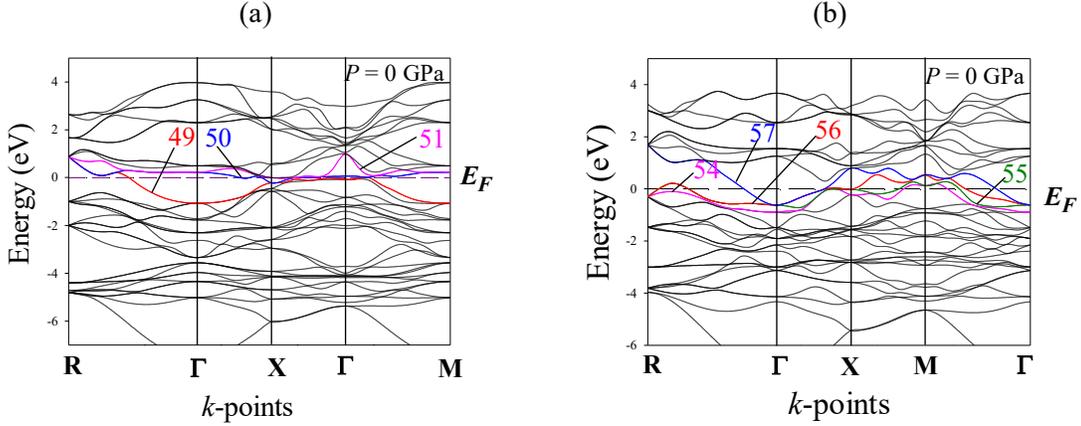

**Figure 4.** Electronic band structure of (a) $Nb_3Pt$ and (b) $Nb_3Os$ along several high symmetry directions of the Brillouin zone at P = 0 GPa (without SOC).

Electronic band structures with SOC for $Nb_3Pt$ and $Nb_3Os$ are shown in Figs. 5 and Figs. 6 respectively. It is found that the band structures of both $Nb_3Pt$ and $Nb_3Os$ split due to SOC. Around the Fermi level, there is a tendency of the bands to become more flatter, for both bonding and antibonding orbital for both the compounds. The band profile of $Nb_3Pt$ and $Nb_3Os$ are very similar to $Nb_3Rh$ and $Nb_3Ir$, respectively [78]. The top of the valence bands of $Nb_3Pt$ and $Nb_3Os$ lies in the energy range from −4.83 eV to Fermi level and −3.95 eV to Fermi level, respectively.

A comparison of band structure of Nb₃Pt and Nb₃Os reveals that the bonding bands are flatter than the antibonding bands around the Fermi level for both compounds because the lower energy band are more localized than higher energy band. These features of band structure have also been pointed out in literature for several A-15 compounds [79-86]. Some common features observed from band structure of A-15 compounds, as has been pointed out before are, around the Fermi level, there is a tendency of the bands to become more flatter, for both bonding and antibonding orbital [30,87,88]. X-ray photoemission spectra studies of the A-15 type compounds have indicated that the Nb $4d$ and the $5d$ energy bands of these A₃B compounds appear to be more and more separate with increasing atomic number of the B element [89-92].

In order to understand the effect of spin on electronic band structure, we have shown the band structure for spin up and spin down electrons of Nb₃Pt and Nb₃Os. Significant effect of spin on band structure is observed because there is splitting of bands both for electrons with up and down spin. For spin up electrons of Nb₃Pt [Fig. 5 (a)], we found that the symmetry has shifted from $\Gamma$-point to $M$-point of the Brillouin zone (compared to non spin-polarized calculation).Figs. 6 show qualitatively similar features for Nb₃Os.

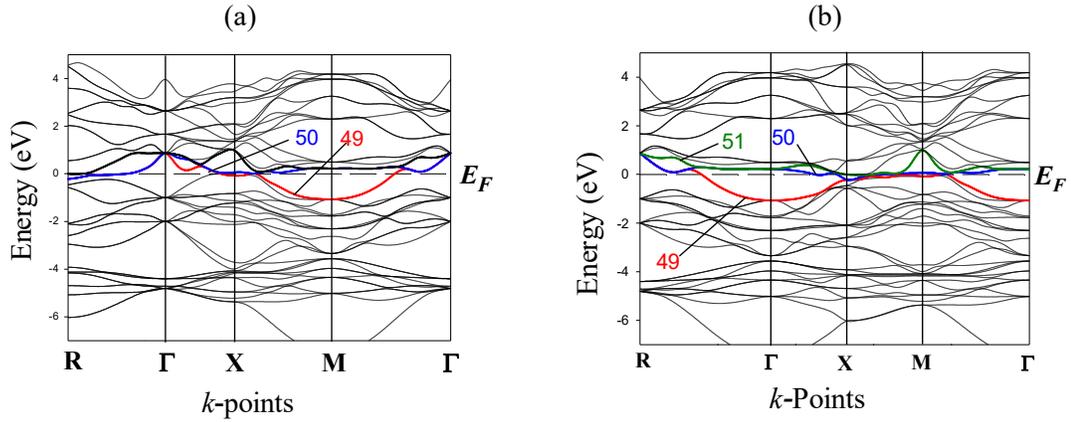

**Figure 5.** Spin-polarized electronic band structure of Nb₃Pt along several high symmetry directions of the Brillouin zone for spin (a) up and (b) down configuration.

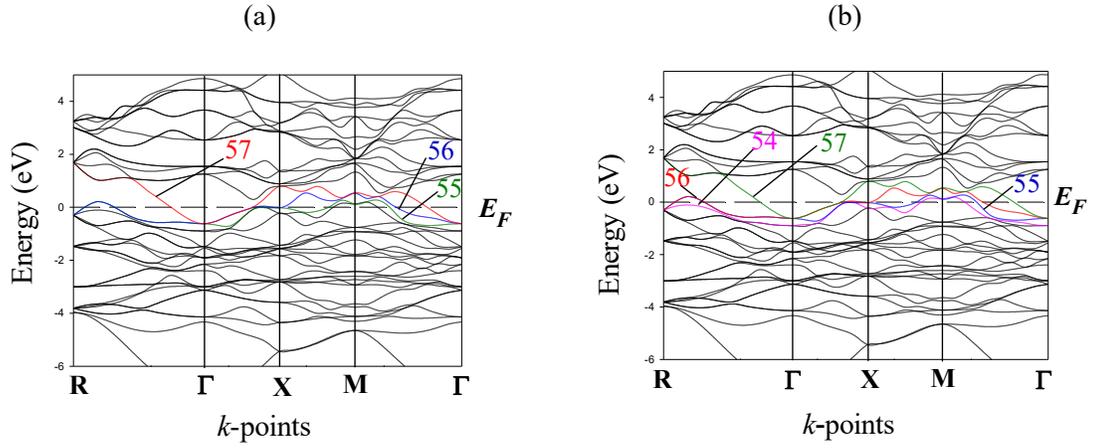

(a)                                    (b)

**Figure 6.** Spin Polarized electronic band structure of Nb₃Os for spin (a) up and (b) down configuration.

### 3.5. *Fermi Surfaces*

It is known that electronic, thermal and some magnetic properties of solids are strongly dependent on the topology of the Fermi surface. In order to understand the electronic transport properties of Nb₃Pt and Nb₃Os, *3D* plots of the Fermi surfaces are shown in Figs. 7 and 8, respectively. The Fermi surfaces are constructed from band numbers 49, 50 and 51 for Nb₃Pt and 54, 55, 56 and 57 for Nb₃Os. The Fermi surfaces of both Nb₃Pt and Nb₃Os contain electron- and hole-like sheets. For band 49 of Nb₃Pt, a hole like topology is found around the symmetry point R. These Fermi sheets are separated from each other. For both the bands 50 and 51 of Nb₃Pt, electron-like sheets are formed around the *X*-point. For band 51, this sheet is small. For band 54 of Nb₃Os, hole-like sheet is seen around the *M*-point. The Fermi surface topology for band 55 of Nb₃Os is quite complex. There is a hole-like sheet around *M*-point and electron-like sheets around *X*- and *R*-points. For band 56 of Nb₃Os, a tiny electron-like sheet appears around the point *R* and electron-like sheet is seen around the central *Γ*-point. For band 57 an electron-like sheet forms around *Γ*-point.

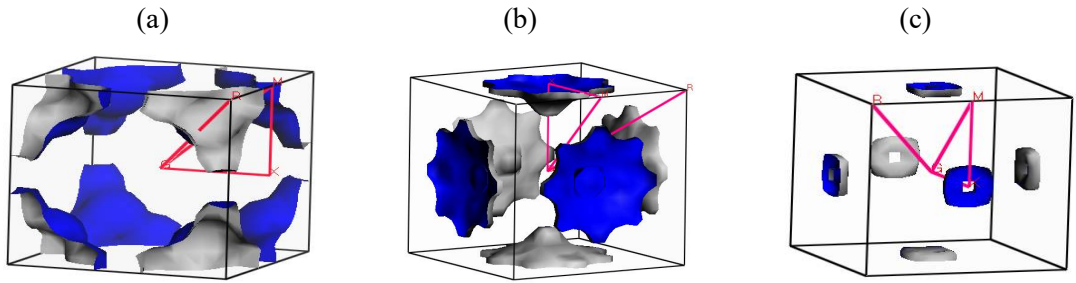

(a)                    (b)                    (c)

**Figure 7.** Fermi surface for bands (a) 49, (b) 50 and (c) 51 of Nb₃Pt, respectively.

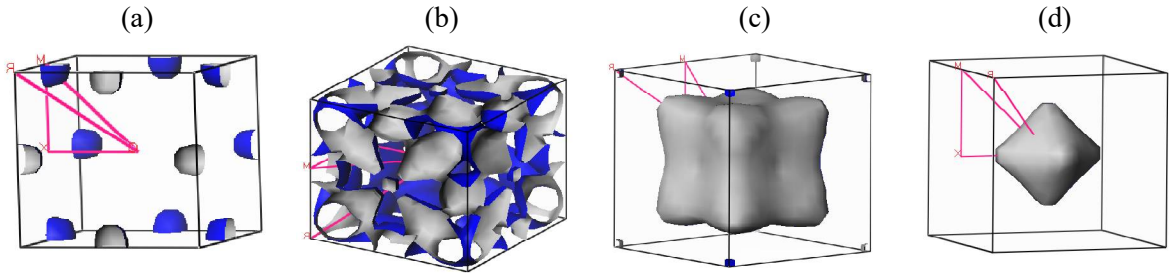

**Figure 8.** Fermi surface for bands (a) 54, (b) 55, (c) 56 and (d) 57 of Nb$_3$Os, respectively.

### 3.6. Electronic charge Density

In order to understand the nature of bonding in Nb$_3$Pt and Nb$_3$Os, we have studied the electronic charge density distribution in the (110) plane (Figs. 9). The charge density distribution map shows that there is an ionic bonding between Nb-Pt and Nb-Os atoms and covalent bonding between Nb-Nb atoms for Nb$_3$Pt and Nb$_3$Os. This agrees with the Mulliken bonding population (Section 3.7).

The color scale on the right side of charge density maps shows the total electron density. The blue color shows the high charge (electron) density and red color shows low charge (electron) density. So, Nb atoms have greater charge density than Pt atoms for Nb$_3$Pt.This is may be due to the *d* orbital of Pt atoms, which is confined in space and whose density of states are quite low. It is also supported by Mulliken bonding population analysis. On the other hand, Os atoms have high charge density compared to Nb atoms for Nb$_3$Os, this also agrees with the Mulliken charge analysis.

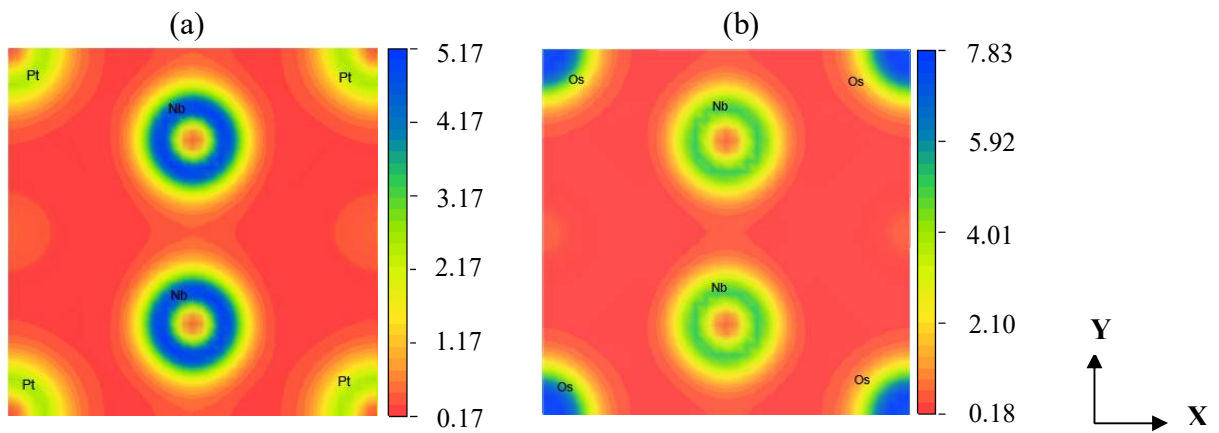

**Figure 9.** The electronic charge density map for (a) Nb$_3$Pt and (b) Nb$_3$Os in the (110).

### 3.7. Bond Population Analysis

To describe the bonding nature further, the Mulliken bond populations [93] are investigated. The results of this analysis are shown in Table 5. The atomic charge of Nb and Pt in $Nb_3Pt$ are 0.21 and -0.62 electron, respectively. Both these values deviated from the normal value expected for a purely ionic state (Nb: +5, +3 and Pt: +4, +2). These deviations partially reflect the covalent character of the bond between Nb atoms (*d-d* hybridized covalent bonding between Nb) and strong iconicity, which is in good agreement with the charge density mapping. Similarly the atomic charge density of Nb and Os in $Nb_3Os$ are 0.15 and -0.46 electron, respectively. Both deviates from the normal value expected in a purely ionic state (Nb: +5, +3 and Os: +4). This deviation again partially reflects the covalent bond character between Nb species that forms by the hybridization between *d* states of Nb, and strong ionicity, which is again in good agreement with the charge density analysis. The charge of Nb in $Nb_3Pt$ is 0.06 electrons larger than that of Nb in $Nb_3Os$. The discrepancy reflects that Nb in $Nb_3Pt$ release more electrons into the conduction band than Nb in $Nb_3Os$. The atomic charge of Os and Pt are -0.46 and -0.62 electron, respectively. The atomic charge of Os is 0.16 electrons larger than that of Pt, which is also indicative of the release of 0.16 more electrons into the conduction band by Os. It is observed that the band spilling parameters are quite low for both the compounds. Which indicates that population analysis presented here is reliable.

For both $Nb_3Pt$ and $Nb_3Os$, electrons are transferred from Nb to Pt and Os, respectively. This suggests that an ionic contribution to the bonding is present. Thus we can say that Nb-Pt and Nb-Os bonding are ionic in $Nb_3Pt$ and $Nb_3Os$, respectively. This is in good agreement with the charge density mapping. The degree of covalency and/or ionicity may be obtained from the effective valence, which is defined as the difference between the formal ionic charge and the Mulliken charge on the cation species [94]. The zero value of effective charge indicates a perfectly ionic bond, while values greater than zero indicate an increasing level of covalency. The effective valence for Nb in $Nb_3Pt$ and $Nb_3Os$ is +2.79 and +2.85 electron, respectively. All these indicate that both ionic and covalent bonds are present in $Nb_3Pt$ and $Nb_3Os$.

Because of the strong basis set dependency of Mulliken Population Analysis (MPA), sometimes MPA gives results in contradiction to chemical intuition. For this reason we have also determined Hirshfeld charge using Hirshfeld Population Analysis (HPA). This has practically no basis set dependency and provides with chemically more meaningful result compared to MPA. Table 6 shows the comparison between Mulliken and Hirshfeld charge. Where we have noted that the magnitude of Hirshfeld charge is in general smaller than that of Mulliken charge.

**Table 5**
Charge spilling parameter (%), orbital charges (electron), atomic Mulliken charges (electron), effective valence (electron) and Hirshfeld charge (electron) in $Nb_3Pt$ and $Nb_3Os$.

| Compounds | Species | Charge spilling | $s$ | $p$ | $d$ | Total | Mulliken charge | Effective valence | Hirshfeld charge |
|---|---|---|---|---|---|---|---|---|---|
| $Nb_3Pt$ | Nb | | 2.30 | 6.52 | 3.97 | 12.79 | 0.21 | +2.79 | 0.02 |
| | Pt | 0.10 | 0.83 | 1.13 | 8.66 | 10.62 | -0.62 | | -0.07 |
| $Nb_3Os$ | Nb | | 2.22 | 6.65 | 3.97 | 12.85 | 0.15 | +2.85 | 0.09 |
| | Os | 0.06 | 2.71 | 7.14 | 6.61 | 16.46 | -0.46 | | -0.28 |

## 4. Conclusions

First principles calculations based on DFT have been used to investigate a number of physical properties of intermetallic $Nb_3Pt$ and $Nb_3Os$ compounds. The equilibrium lattice parameters at zero pressure of these compounds are in very good agreement with both theoretical and experimental values. The calculated elastic parameters allow us to conclude that both the compounds are mechanically stable. Young's modulus of $Nb_3Os$ is larger than $Nb_3Pt$, which indicates that $Nb_3Os$ is much stiffer than $Nb_3Pt$. Poisson's ratio, Cauchy pressure, and Pugh's ratio indicate that $Nb_3Pt$ and $Nb_3Os$ should be ductile metallic material. The calculated shear anisotropy factors assure that both the compounds are elastically anisotropic. From $B/C_{44}$ values of gold, we can conclude that $Nb_3Os$ is expected to have good lubricating properties compared to $Nb_3Pt$. The Debye temperatures of the intermetallics are low. The calculated values of $\theta_D$ using the estimated elastic moduli agree reasonably well with experimental values [72]. The electronic energy band structure and total density of states analyses reveal that both the materials are metallic in nature, where the conductivity is expected to be dominated by the Nb-4$d$ electronic orbitals. Considering the value of $N(E_F)$, electrical conductivity of $Nb_3Pt$ is expected to be higher than that of $Nb_3Os$. Band structure calculations were performed with and without SOC. The degree of spin anisotropy in the TDOSs is very small for both the compounds. DOS features indicate that $Nb_3Os$ is electronically more stable than $Nb_3Pt$. Fermi surface of both the intermetallics exhibits a complex combination of electron and hole like sheets. Result of Mulliken population analysis and charge density mapping suggest that $Nb_3Pt$ and $Nb_3Os$ have both ionic and covalent bonding in addition to metallic bonding. For both $Nb_3Pt$ and $Nb_3Os$, electrons are transferred from Nb to Pt or Os. Maximum charge density is found around Nb (for $Nb_3Pt$) and Os (for $Nb_3Os$) atoms, respectively. Mulliken population analysis also shows that Nb-Nb bonding is stronger than Nb-Pt/Nb-Os bonding in $Nb_3Pt$ and $Nb_3Os$, respectively. Nb-Nb bonds of $Nb_3Pt$ are more covalent in nature than that in $Nb_3Os$.

To conclude, this study presents a detailed DFT based *ab-initio* study of elastic, electronic, and bonding properties of technologically important intermetallic compounds $Nb_3Pt$ and $Nb_3Os$, for the first time. We hope that this work will inspire other groups to study these interesting materials further.